\documentclass[twocolumn]{revtex4}

\usepackage{graphicx}
\usepackage{t1enc}
\usepackage{times}

\begin{document}

\title{Phase field theory of crystal nucleation in hard sphere liquid}

\author{László Gránásy}
\author{Tamás Pusztai}
\author{Zoltán Jurek}
  \affiliation{Research Institute for Solid State Physics and Optics,
               H--1525 Budapest, POB 49, Hungary}
\author{Massimo Conti}
  \affiliation{Dipartimento di Matematica e Fisica, Universita' di
               Camerino, and Istituto Nazionale di Fisica della Materia,
               Via Madonna delle Carceri, I-62032, Camerino, Italy}
\author{Bj{\o}rn Kvamme}
  \affiliation{University of Bergen, Department of Physics,
               All\'egaten 55, N--5007 Bergen, Norway}

\date{\today}

\begin{abstract}
The phase field theory of crystal nucleation described in
[L.~Gr\'an\'asy, T.~B\"orzs\"onyi, T.~Pusztai, Phys.\ Rev.\
Lett.\ {\bf 88}, 206105 (2002)] is applied for nucleation
in hard--sphere liquids. The exact thermodynamics from molecular
dynamics is used. The interface thickness for phase field is
evaluated from the cross--interfacial variation of the height of the
singlet density peaks. The model parameters are fixed in
equilibrium so that the free energy and thickness of the
(111), (110), and (100) interfaces from molecular dynamics are
recovered. The density profiles predicted without adjustable
parameters are in a good agreement with the filtered densities
from the simulations. Assuming spherical symmetry, we evaluate
the height of the nucleation barrier and the Tolman length
without adjustable parameters. The barrier
heights calculated with the properties of the (111) and (110)
interfaces envelope the Monte Carlo results, while those obtained with
the average interface properties fall very close to the exact values.
In contrast, the classical sharp interface model considerably
underestimates the height of the nucleation barrier. We find that
the Tolman length is positive for small clusters and decreases with
increasing size, a trend consistent with
computer simulations.
\end{abstract}

\maketitle
\section{Introduction}
In this paper we present a detailed quantitative test of the phase field
theory of crystal nucleation in the hard sphere (HS) fluid, a case in
which all the data necessary for such analysis are available.

Probably the least understood stage of the freezing of liquids
is the initial nucleation phase in which the crystalline phase appears
in the homogeneous liquid via heterophase fluctuations whose
central part shows crystal-like
atomic arrangement. Those heterophase fluctuations that exceed a
critical size (determined by the interplay of the interfacial and
volumetric contributions to the cluster free energy) have a good
chance to reach macroscopic dimensions, while the smaller ones
decay with a high probability. The description of such fluctuations
is problematic even in single component systems. One of the main
difficulties is that the typical size of the critical fluctuations
that form on the human time scale \cite{SA90,BC95,B98,WF95,AF01a} is
comparable to the thickness of the crystal-liquid interface, which
in turn extends to several molecular layers \cite{BG86,LH92,DL98}.
Therefore, the droplet model of the classical nucleation theory (CNT)
that relies on a sharp interface
and bulk crystal properties is expected to be inappropriate for
describing such fluctuations. Field theoretic models, that predict
a diffuse interface, offer a natural way to handle such problems
\cite{O02}. For example, in a recent paper \cite{GBP02}, the phase
field theory (PFT) has been shown to describe such fluctuations
quantitatively.

Due to extensive computer simulations done recently, the
hard sphere (HS) liquid is probably the best known system that
shows crystal nucleation. Its thermodynamic properties can be
obtained accurately by integrating the equations of state of
the crystalline and liquid phases evaluated from molecular
dynamics (MD) \cite{H72}. The interface density
profiles are known for the (111) and (100) interfaces \cite{DL98}.
The free energy of the (111), (110), and (100) interfaces has
been evaluated with a high accuracy \cite{DL00}. Furthermore,
the height of the nucleation barrier has been determined
by Monte Carlo (MC) simulations \cite{AF01b}. This
offers a unique possibility to test the abilities of various
cluster models. For example, it has been established that the
droplet model of the classical nucleation theory (that relies
on the equilibrium value of the interface free energy) seriously
underestimates the height of the nucleation barrier, and the
agreement could only be restored via assuming
that the interface free energy increases with
supersaturation \cite{AF01a,AF01b}. Such behavior has been
recovered by a single-order-parameter density functional
approach based on a Ginzburg-Landau free energy consistent
with face centered cubic (fcc) crystal symmetries \cite{GP02}.
Remarkably, however, the interface profiles from models
deduced for fcc symmetries \cite{SO96a,SO96b,GP02} do not
fit well to the simulation results. It is, therefore,
desirable to refine the field theoretic models so
that details of the interface profiles are reproduced
together with the height of the nucleation barrier. A possible
starting point for such study is the phase field theory,
that has been used successfully to describe many aspects
of crystallization \cite{BWBK02,GPWDF03}, including
nucleation \cite{GBP02}.

Besides these, further interest in describing crystal nucleation
in the hard sphere system is
generated by recent experiments on colloidal suspensions that
mimic closely the hard-sphere behavior \cite{HS}.

Herein, the hard sphere system is used to test the performance of the phase
field theory. The model parameters of the phase field theory are fixed
at the solid--liquid equilibrium so that the known interface thickness
and free energy data are recovered. Then, the nucleation
barrier height is predicted without adjustable parameters, which is
compared to exact results from Monte Carlo simulations.
It will be shown that the phase field theory leads
to a considerable improvement relative to the classical droplet
model. Having determined the interface density profiles, we
evaluate the Tolman length, a quantity related to the curvature
dependence of the interface free energy. We find that in agreement
with computer simulations, the Tolman length is positive for small
clusters and decreases towards a negative value with increasing 
cluster size.

\section{Nucleation theory}

\subsection{Phase field theory of nucleation}

In the present work, we apply two fields to describe the local state of matter:
A non-conserved structural field $m$ called phase field, and the
conserved volume fraction field $\phi$, which is related to the
density field as $\phi=\frac{\pi}{6} \sigma^3 \rho$, where
$\sigma$ is the HS diameter, and $\rho$ is the number density.
For historical reasons, $m$ is defined so that it is 1 for the liquid
and 0 for the solid. Then, $1-m$ can be regarded as a local degree of
crystallinity, and can be viewed as the Fourier amplitude of the
dominant density wave in the Fourier expansion of the time averaged
number density in the crystal.

In line with the formulation for binary systems \cite {WB95,CJ95,GBP02,GPWDF03},
the Helmholtz
free energy of the inhomogeneous hard sphere system is assumed to have the form
\begin{equation}
  F = \int d{\bf r} \bigg\lbrace \frac{bT}{2}(\nabla m)^2 + f(m,\phi)
  \bigg\rbrace,
\end{equation}
\noindent
where the local Helmholtz free energy density is given as
\begin{equation}
f(m,\phi) =
wTg(m) + [1-p(m)] f_S(\phi) + p(m) f_L(\phi),
\end{equation}
\noindent
the $g$ ``double well''
and $p$ ``interpolation'' functions have the usual form
$g(m)=\frac{1}{4} m^2 (1-m)^2$ and $p(m) = m^3(10-15m+6m^2)$,
emerging from the thermodynamically consistent formulation of
the phase field theory \cite{WEA93,WB95}, while $f_S(\phi)$ and $f_L(\phi)$
are the free energy densities of the bulk crystal and liquid phases.
The square gradient (SG) term in the integrand is responsible for
the diffuse interface.
Being in unstable equilibrium, the critical fluctuation (the nucleus)
can be found as an extremum of this free energy functional. Then the
fields have to obey the appropriate Euler-Lagrange (EL) equations, which
in the case of such local functional take the form 
\begin{equation}
\frac{\delta F}{\delta \chi} =
\frac{\partial \psi}{\partial \chi}
-\nabla \frac{\partial \psi}{\partial \nabla \chi} = 0,
\end{equation}
\noindent
where $\frac{\delta F}{\delta \chi}$ is the first functional derivative 
of the free energy with respect to the field $\chi(= m$ or $\phi)$, 
and $\psi$ is the total local free energy density [the integrand of 
Eq.~(1)]. These equations have to be solved under the boundary conditions
of having unperturbed liquid far from the fluctuation, and due to 
symmetry reasons, zero field gradients at the center of the fluctuations.
Note, that since the volume fraction is a conserved field,
we have to add
$\lambda \int d{\bf r} \phi= -\mu_{\infty} \int d{\bf r} \rho$ to the right 
hand side of Eq.~(1), when searching for the extremum of the free energy,
where the Lagrange multiplicator $\lambda$ turns out
to be proportional to the negative of the chemical 
potential $\mu_{\infty}$ of the unperturbed
liquid. This is then the Legendre transformation that yields the
grand potential, $\Omega = F - \mu_{\infty} N$, i.e., we seek the extremum
of the grand potential functional. Assuming spherical symmetry,
the EL equations boil down to the following ones
\begin{equation}
\frac{\delta \Omega}{\delta m}= \frac{\partial \psi}{\partial m} 
- bT \lbrace m''+ \frac{2}{r} m' \rbrace = 0
\end{equation}
and
\begin{equation}
\frac{\delta \Omega}{\delta \phi}= \frac{\partial \psi}{\partial \phi}
-\frac{\partial \psi}{\partial \phi}(r \rightarrow \infty) = 0,
\end{equation}
\noindent
where $r \rightarrow \infty$ means
that the term has to be evaluated
far from the fluctuation. [In this work ' stands for differentiating 
with respect to the argument where argument is shown; otherwise 
(as above), it 
denotes differentiation with respect to the radial distance $r$ from
the center of the fluctuation.]
Note that Eq.~(5) is an implicit equation for
the volume fraction $\phi = \phi(m)$,  
\begin{equation}
0=[1-p(m)] \frac{\partial f_S}{\partial \phi}
+p(m)\frac{\partial f_L}{\partial \phi}
- \frac{\partial f_L}{\partial \phi}(r \rightarrow \infty)
\end{equation}
\noindent
and thus
\begin{equation}
\frac{\partial \psi}{\partial \phi} = w T g'(m) + p'(m)[f_L(\phi)-f_S(\phi)] 
\end{equation}
\noindent
is a function of $m$, i.e., Eq.~(4) is an ordinary differential
equation for $m$.

The properties of the critical fluctuation can be determined by 
solving Eq.~(4) under the boundary conditions
$m' \rightarrow 0$ for $r \rightarrow 0$, and $m \rightarrow 1$ 
and $\phi \rightarrow \phi_{\infty}$ for $r \rightarrow \infty$
that correspond to the initial non-equilibrium liquid state.
In this work, Eq.~(4) has been solved numerically, using a
variable 4th/5th order Runge-Kutta method. Since the boundary 
conditions fix $m$ and $m'$ at different positions, the phase field
value at the center, $m(r \rightarrow 0)$, 
for which the far-field condition is satisfied has been found 
iteratively.

Having determined $m(r)$, the volume fraction/density profile can
be obtained by inserting it into the numerically inverted
version of Eq.~(6). The free energy of the critical fluctuation,
in turn, can be calculated by inserting the solution $m(r)$ into
\begin{equation}
W^* = \int_0^\infty dr 4\pi r^2 \Delta \omega[m,\phi(m)] , 
\end{equation}
\noindent
where $\Delta \omega = \omega - \omega_{\infty}$, and the grand 
potential density is $\omega = \psi - \mu_{\infty} \rho$,
while $\omega_{\infty} = \psi_{\infty} - \mu_{\infty} \rho_{\infty}$. 
Here, subscript ${\infty}$ denotes properties of the initial liquid. 

The interface free energy of small fluctuations is expected to depend
on size or curvature due the reduction of the average number of solid 
neighbors of surface molecules relative to the planar interface.
The analogous phenomenon in small liquid droplets has been studied 
extensively \cite{RW82}. In the widely acknowledged thermodynamic 
theory of Tolman \cite{T49}, the size dependence of the surface 
tension is given in terms of  the Tolman length \cite{DEF} $\delta_T =
R_e - R_p$ (the distance of the equimolar surface \cite{DEF1} from the
surface of tension \cite{DEF2}) as
\begin{equation}
\gamma=\frac{\gamma_{\infty}}{1 + 2\delta_T/R_p},
\end{equation}
\noindent
where $\gamma_{\infty}$ is the surface tension for planar geometry. 
Although a rigorous derivation of these notions is unavailable for 
crystallites, a quantity analogous to $\delta_T$ has recently been 
evaluated from computer 
simulations \cite{BC95}. It decreases with increasing size of the 
fluctuations. It is of considerable interest to see whether the 
phase field theory is able to reproduce this feature.

We determine the Tolman length here via three routes: 

(i) Using the radius of the equimolar surface,
$R_e = \lbrace\int_{0}^{\infty} 
dr 4\pi r^2 [\phi-\phi_{\infty}]/
[(4\pi/3)(\phi_0-\phi_{\infty})]\rbrace^{1/3}$, where $\phi_0$ is the
value at the center of the fluctuation, and the
expression $R_p = [3W^*/(2\pi \Delta \omega_{max})]^{1/3}$ \cite{relat} 
for the radius of the surface of tension. Here 
$\Delta \omega_{max} = max \lbrace f_L(\phi_{\infty})+
(\partial f_L/\partial \phi)_{\phi_{\infty}}(\phi-\phi_{\infty})
-f_S(\phi)\rbrace$ is the maximum driving force of crystallization,
realized by the (compressed) solid that has the same chemical 
potential as the initial liquid \cite{CH59}.
  
(ii) On the basis of Eq.~(9),
 $\delta_{T,eff} = \frac{1}{2}R_p (\gamma_{\infty}/\gamma -1)$, 
where the ratio of the curved and planar interfaces is
related to the ration of the non-classical and classical cluster
free energies as $\gamma/\gamma_{\infty} = (W^*/W^*_{CNT})^{1/3}$
\cite{relat}.

(iii) On the basis of Eq.~(9), but approximating the radius of the
surface of tension with that of the equimolar surface,
 $\delta_{T,eff} = \frac{1}{2}R_e (\gamma_{\infty}/\gamma -1)$, as
done in the evaluation from MD results \cite{BC95}.

Note that Eq.~(9) has been derived from the hypothesis of a
size-independent Tolman length, a condition that is 
not satisfied in the known cases \cite{Tolm}, thus routes (ii) 
and (iii) are expected to be less accurate.

Provided that $f_L(\phi)$ and $f_S(\phi)$ are known, the phase field theory 
contains only two parameters, the coefficient of the square-gradient term
$b$ and the free energy scale $w$. It is worth recalling in this respect 
that these quantities can be related to the free energy and thickness 
of the equilibrium planar interfaces \cite{CH59}.
\begin{equation}
\gamma_{\infty} = (2bT)^{1/2} \int_{0}^{1} d\xi f[\xi,\phi(\xi)]^{1/2},
\end{equation}
\noindent
and 
\begin{equation}
d_{10-90} = (\frac{bT}{2})^{1/2} \int_{0.1}^{0.9} d\xi f[\xi,\phi(\xi)]^{-1/2}, 
\end{equation}
\noindent
where $d_{10-90}$ is the 10$\%$ -- 90$\%$ thickness of the interface, the
distance on which $m$ varies between 0.1 and 0.9. Thus, if 
$\gamma_{\infty}$ and 
$d_{10-90}$ are known, these relationships can be used to fix $b$ 
and $w$ in equilibrium. This then allows the calculation of the free
energy of the critical fluctuations $W^*$ and the Tolman length
{\it without adjustable parameters}.

\subsection{Classical theory}

For comparison with our non--classical approach, we calculate 
the height of the nucleation barrier using the sharp interface
droplet model of the classical nucleation theory \cite{K91}. 
In this approach, the free energy of heterophase fluctuations of 
radius $R$ is given as $W_{CNT}= -(4\pi/3) R^3 \Delta \omega_{max} +
4 \pi R^2 \gamma_{\infty}$. Then, the free energy of the critical 
fluctuation of radius $R^*_{CNT}=2 \gamma_{\infty}/\Delta \omega_{max}$ 
reads as $W^*_{\rm CNT} = (16\pi/3)\gamma_{\infty}^3\Delta 
\omega_{max}^{-2}$.

\subsection{Application to hard spheres} 

The stable crystalline phase of the hard sphere system has the
face centered cubic (fcc) structure. The nuclei, in turn, contain 
fcc (ABC) and hexagonal close packed (hcp) stacking (ABA) with
equal probability \cite{AF01a},
indicating that the nucleation barrier for them is rather similar.
Therefore, we address here only fcc nucleation.
 
The Gibbs free energy difference between the fcc solid and liquid 
(the driving force of crystallization) has been obtained by integrating 
the virtually accurate equations of state by Hall~\cite{H72} (that fit
to the molecular dynamics results), starting from the 
volume fractions of the coexisting solid and liquid $\phi_{{\rm fcc},{\rm
coex}} = 0.546$ and $\phi_{{\rm L},{\rm coex}} = 0.494$ at pressure $p
= 11.81 kT/\sigma^3$. Polynomials were then fitted to the chemical potentials 
of the solid and liquid phases, whose relative deviation from the 
accurate chemical potentials are less then $10^{-4}$ in the density 
range of interest. These polynomials were then integrated analytically 
to obtain the free energy densities of the two bulk phases.

Since the structural order parameter $1-m$ can be interpreted 
as the Fourier 
amplitude of the dominant density waves (with the first neighbor 
reciprocal lattice vectors as the wave number), we identify the 
phase field profile as the cross-interfacial variation of the
height of the density peaks. This is an approximation,
since other Fourier components also contribute. 
However, as pointed out by Shen and Oxtoby \cite{SO96a},
so far as the density of solid is reasonably approximated 
by Gaussians centered around the fcc lattice sites, the
amplitudes of all other components are proportional to the 
amplitude of the dominant wave, i.e., to a first approximation
the height of the density peaks is proportional to the amplitude
of the dominant Fourier component.
 
The interfacial density profiles of the equilibrium hard-sphere
crystal-liquid interfaces of orientations (100) and (111) have been 
studied by Davidchack and Laird using molecular dynamics 
simulations~\cite{DL98}. We have received new high resolution 
data from them for the (111), (100), and (110) interfaces. 
In the latter cases, the bin size for the density profile was
1/32 times the layer width, while the parameters for Gaussian
filtering were $N = 64$ and $\epsilon = 28.1$ \cite{D03}.
(For definitions see \cite{DL98}.) The new density distributions 
are fully consistent with those in~\cite{DL98}.
From the interfacial variation of the peak amplitudes, we obtained
5.94$\sigma$, 4.75$\sigma$, and 5.54$\sigma$
 for the 10$\%$--90$\%$ interface thickness for the (111), (110),
and (100) directions, respectively.

The same authors evaluated the free energy of the fcc crystal --
liquid interface for these orientations \cite{DL00}: $\gamma_{\infty}
\sigma^2/kT = 0.58 \pm 0.01$, $0.62 \pm 0.01$, and $0.64 \pm 0.01$
for the $(111)$, $(100)$ and $(110)$ interfaces.
Considering that due to the small anisotropy the equilibrium shape
is expected to be nearly spherical, we also performed calculations
with an average of the interface free energy over the known orientations
$\gamma_{\infty} \sigma^2/kT = 0.6133$.

Having this information available, the model parameters $b$ and $w$
were determined using Eqs.~(10) and (11) via Newton-Raphson
iteration. The Helmholtz free energy density [Eq.~(2)] obtained with $w$
evaluated using the properties of the (111) interface is shown in
Fig.~1. Qualitatively similar free energy surfaces were obtained for
the other orientations.

\begin{figure}
\includegraphics[width=\columnwidth]{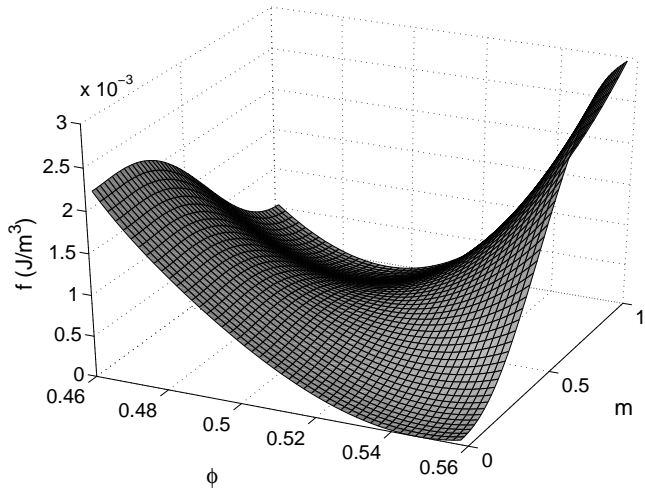}
\caption{Free energy surface for the hard sphere system evaluated
with free energy scale $w$ corresponding to the (111) interface.
Note, that the minima at $(\phi=0.494, m=1)$ and $(\phi=0.546,
m=0)$ correspond to the equilibrium liquid and solid (fcc) phases.
The calculation has been performed for a colloidal suspension of
hard sphere diameter $\sigma =$ 890 nm, a particle size comparable
to that in many experiments \cite{HS}. The small value of the free
energy density follows from this large molecular diameter.
}
\end{figure}

Using these data, the PFT and CNT predictions for the nucleation
barrier can be made free of adjustable parameters.

\section{Results and discussion}

\subsection{Density profiles}

Fixing the model parameters $b$ and $w$ as described above, the
phase field profile for the equilibrium solid-liquid
interface can be determined via the equation \cite{CH59}
\begin{equation}
z(m) = (\frac{bT}{2})^{1/2} \int_{0.5}^{m}
d\xi f[ \xi,\phi(\xi)]^{-1/2},
\end{equation}
\noindent
where $z$ is the distance perpendicular to the plane $m = 0.5$.
The coarse grained density profile that corresponds to the filtered
density by Davidchack and Laird \cite{DL98} can be obtained by
inserting the solution $m(z)$ into Eq.~(6) after inverting it
numerically.

The predicted structural order parameter $(1-m)$ and normalized
coarse grained density $(\rho-\rho_{min})/(\rho_{max}-
\rho_{min})=(\phi-\phi_{min})/(\phi_{max}-
\phi_{min})$ profiles are compared with their counterparts
from molecular dynamics simulations in Figs.~2 to 4. For the sake
of comparison, the simulation results are also presented in a normalized
form, $(X-X_{min})/(X_{max}-X_{min})$, where $X$ is either
the density peak height or the filtered density, while $X_{min}$ and
$X_{max}$ are the minimum and maximum values of these quantities.

\begin{figure}
\includegraphics[width=\columnwidth]{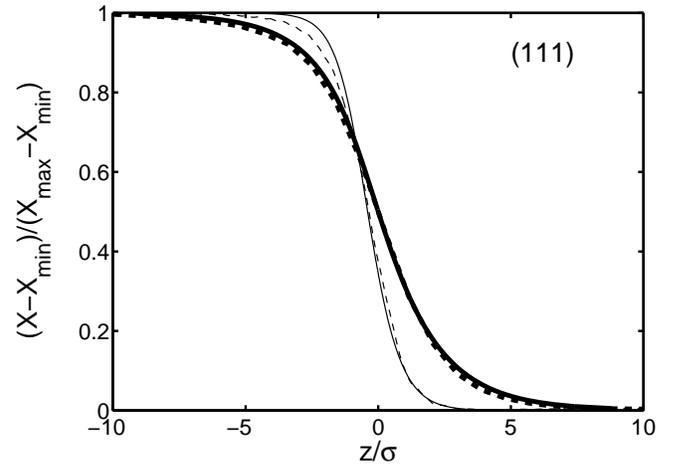}
\caption{Reduced interfacial profiles for the (111) interface.
(Solid lines: predictions of the phase field theory; heavy solid line
-- structural order parameter, $1-m$; light solid line -- density. 
Dashed lines: Results from
molecular dynamics \cite{DL98}; heavy dashed line -- height of singlet density
peaks; light dashed line -- filtered density profile. Note that the heavy
lines should be compared to each other, as the light ones.)}
\end{figure}

\begin{figure}
\includegraphics[width=\columnwidth]{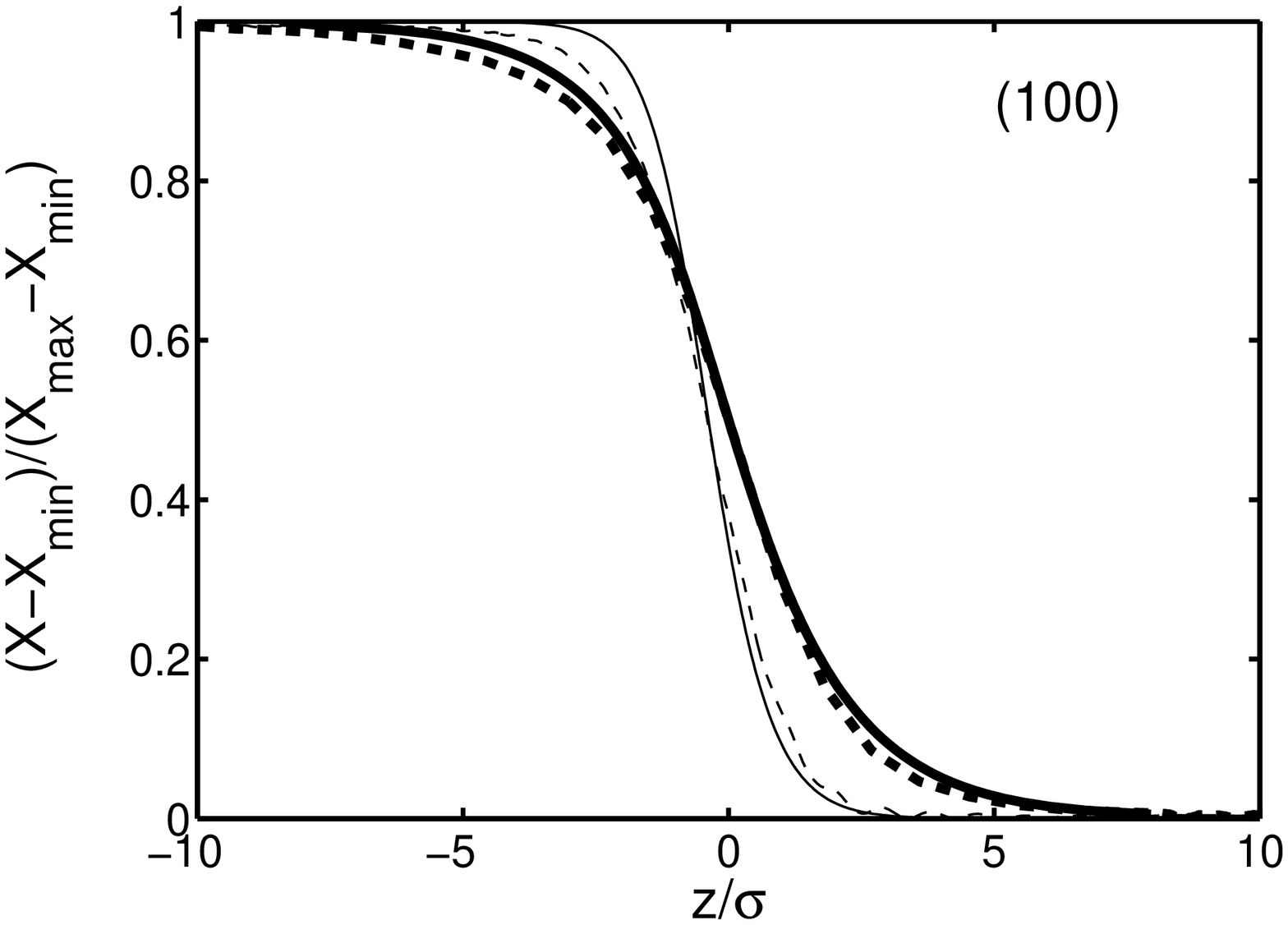}
\caption{Reduced interfacial profiles for the (100) interface.
Notation as for Fig.~2.}
\end{figure}

\begin{figure}
\includegraphics[width=\columnwidth]{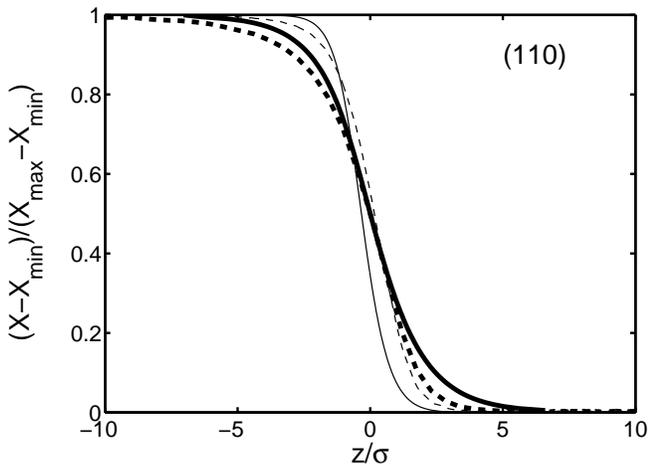}
\caption{Reduced interfacial profiles for the (110) interface.
Notation as for Fig.~2.}
\end{figure}

For the (111) and (100) interfaces, and to a smaller extent for the
(110) interface, the structural order parameter profiles are in a good
agreement with the cross-interfacial variation of the density peak height
from MD. Note that while $d_{10-90}$ has been fitted, the agreement
depends also on the shape/symmetry of the predicted profile. It appears,
that the nearly symmetric structural order parameter profile of the PFT
approximates reasonably well the behavior seen in the simulations.
Remarkably, the density functional theories by
Shen and Oxtoby \cite{SO96a,SO96b},
and Gr\'an\'asy and Pusztai \cite{GP02} derived for the fcc
structure predict significantly more asymmetric structural order
parameter profiles for the crystal--liquid
interfaces, which are sharper on the crystal side and have a long tail
on the liquid side. Therefore, their match to the simulated profiles
is apparently less
satisfactory. It is an intriguing question why the quartic form of
$g(m)$, that is consistent with the symmetries of the
base centered cubic (bcc) structure \cite{SWZS87},
leads to a better fit to the simulation profile then those obtained
by using free energy functionals consistent with the fcc symmetries.
A distinct possibility is the presence of a bcc-like layer at the
interface as reported in the case of the Lennard-Jones system
\cite{WF95,SO96a}.
This question, however, could only be addressed in a full density
functional theory (such as that by Shen and Oxtoby \cite{SO96c})
that incorporates the possibility for the Bain distortion that
connects the bcc and fcc structures.

The coarse grained density profiles can now be predicted in the
PFT without adjustable parameters. They appear to be in a good agreement
with the simulation results for the (111) and (100) interfaces (Figs.~2
and 3), however, a shift of about half a $\sigma$ can be seen between
them for the (110) direction, the PFT result lying closer
to the crystal (Fig.~4).

Summarizing, in the PFT, we have a free energy surface and a SG
coefficient that reproduce exactly the thermodynamics of the bulk
phases, the interface free energy, and quite reasonably both the
structural order parameter profile and the density profile. Thus,
we may attempt to calculate the height of the nucleation barrier
with some confidence.

\subsection{Nucleation}

The reduced nucleation barrier heights calculated with $b$ and $w$
that fit to the properties of the (111) and (110) interfaces and
to the average interface properties are shown in Fig.~5 for the
phase field theory. For comparison, the predictions of the classical
nucleation theory and the exact results from Monte Carlo 
simulations~\cite{AF01b} are also presented. The $W^*$ vs. initial 
liquid volume fraction curves predicted by the PFT
using the properties of the (111) and (110) interfaces envelope the
MC simulations, while the predictions with the average interface
properties fall close to the MC results. The slope of the $W^*$ vs.
$\phi_{\infty}$ curve appears to be somewhat larger for the PFT
predictions than for the MC simulations. A possible explanation
could be the density dependence of the parameters $b$ and $w$
\cite{GP02}. However, the more complex Euler-Lagrange equations
associated with such problem will be investigated elsewhere.

\begin{figure}
\includegraphics[width=\columnwidth]{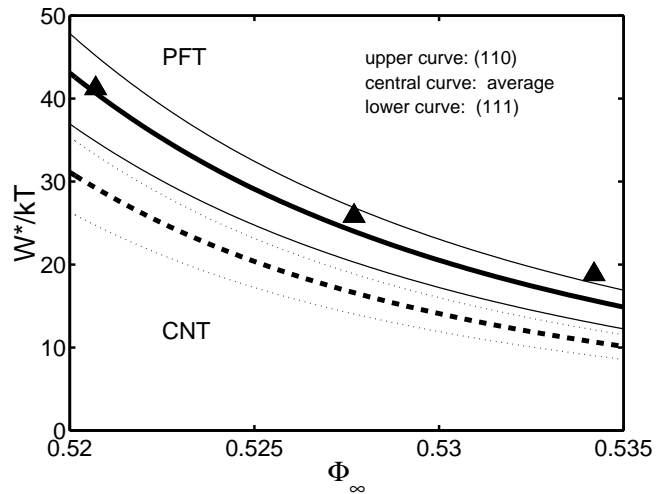}
\caption{Reduced nucleation barrier height vs. volume fraction
of the initial liquid. (Solid lines: phase field theory -- PFT;
dashed/dotted lines: classical droplet model -- CNT.)
The upper an lower curves were calculated using the physical
properties of the (111) and (110) interfaces, respectively.
The prediction for the (100) interface (not shown here) falls
slightly below the results obtained with the average interface
properties (heavy lines). For comparison, the results
of Monte Carlo simulations \cite{AF01b} are also shown (triangles).}
\end{figure}

It is of interest to determine the magnitude of the error associated with
these results. While the relevant thermodynamic data are
virtually exact, uncertainties
of $(\pm 0.01 kT/\sigma^2)$ and $(\pm 5 \%)$ are associated with
the interface free energy and the interface thickness.
We find that for $\phi = 0.5207$ the barrier height is
$W^*/kT = 40.63$, while its uncertainties originating from those
of the interface free energy and the interface thickness amount
to $\pm 1.91$ and $\pm 0.23$, respectively, yielding $\pm 1.92$
when assuming Gaussian error propagation. The magnitude of these errors
varies approximately proportionally with the barrier height. These
errors do not influence the validity of our conclusions.

The radial order parameter and density profiles corresponding
to the volume fractions used in MC simulations (0.5207,
0.5277, and 0.5343) are shown in Fig.~6. The interfacial profiles
for $\phi_{\infty}=0.5207$ are in
a reasonable agreement with the size of the respective critical
fluctuation from MC simulation ({\it cf.} our Fig.~6 and Fig.~3 of
Ref.~\cite{AF01a}). Note that the bulk crystalline properties have not
been established even at the centers of the critical fluctuations.
Accordingly, the classical droplet model
significantly underestimates the height of the nucleation
barrier for all orientations. Since the nucleation rate is
proportional to $exp(-W^*/kT)$, these differences amount to
orders of magnitude. For example, the calculation made
with the average interface free energy overestimates
the nucleation rate by three to five orders of magnitude.
At the volume fraction $\phi_{\infty} = 0.5781$,
the PFT and CNT predictions intersect each other. In both
approaches, the height of the nucleation barrier decreases
monotonically with the supersaturation.

\begin{figure}
\includegraphics[width=\columnwidth]{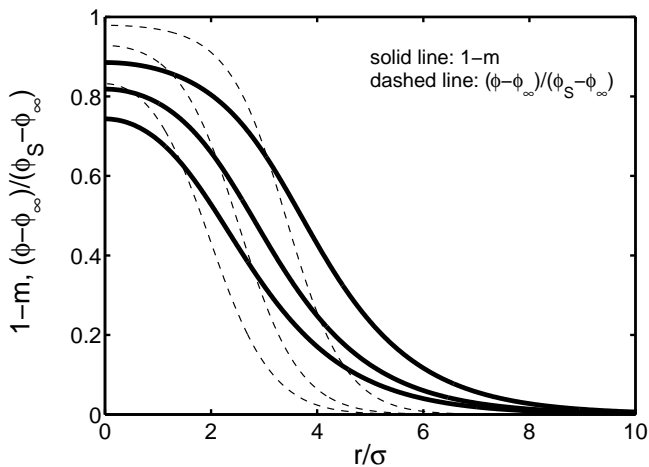}
\caption{Radial structural order parameter (solid lines) profiles and
reduced density (dashed lines) profiles at initial liquid volume fractions
$\phi_{\infty} = 0.5207, 0.5277$, and 0.5343, calculated with
$b$ and $w$ corresponding to the average interface properties
($\gamma_{\infty} \sigma^2/kT=0.6133$ and $d_{10-90}/\sigma = 5.410$).
Here $\phi_S (= 0.5796, 0.5875,$ and 0.5948, respectively) is the volume
fraction of the solid that is in mechanical equilibrium with the
initial liquid (the crystal that has the same pressure
as the liquid of volume fraction $\phi_{\infty}$).
}
\end{figure}

The three known orientations can in principle be used to approximate
the anisotropy of the interface free energy \cite{DL03,HAK01}, that
in turn defines the equilibrium shape of minimum interface
free energy for given volume. Since the nuclei are expected to minimize
their free energy, their {\it average} shape can
probably be reasonably approximated by the equilibrium form. Work is
underway to determine the nucleation barrier in such case.

Finally, we draw attention to the fact, that the MC data for
$W^*/kT$ \cite{AF01a,AF01b}, used here as reference, were obtained directly
via ``umbrella sampling'', therefore, possible uncertainties
associated with the nucleation prefactor~\cite{B98} do not
plague them.

\subsection{Tolman length}

The Tolman lengths calculated for the critical fluctuations
along the three routes specified above are shown as a function
of the initial volume fraction in Fig.~7. Similarly to the field
theoretic results for vapor-liquid nucleation \cite{Tolm}, 
for small sizes the Tolman length is positive and comparable 
to the molecular diameter, while $\delta_T$ decreases 
with increasing size towards a negative limiting value for the 
planar interface. Although this behavior is general for the Tolman
lengths from all three routes, routes (ii) and (iii) yield
considerably smaller values than the more accurate route (i).
Accordingly, the change of sign
happens at different volume fractions: At $\phi_{\infty}=0.5266$
for route (i), while at $\phi_{\infty} = 0.5781$ for routes
(ii) and (iii). The differences originate from the fact that
the assumption $\delta_T=const.$, routes (ii) and (iii) rely on,
is not satisfied for small clusters. In turn, in the large particle
limit this assumption becomes valid. Indeed, the Tolman lengths
from all three routes converge to the same negative
value $({\delta_{T,eq}}/{\sigma}=-0.33)$ in the planar limit.

\begin{figure}
\includegraphics[width=\columnwidth]{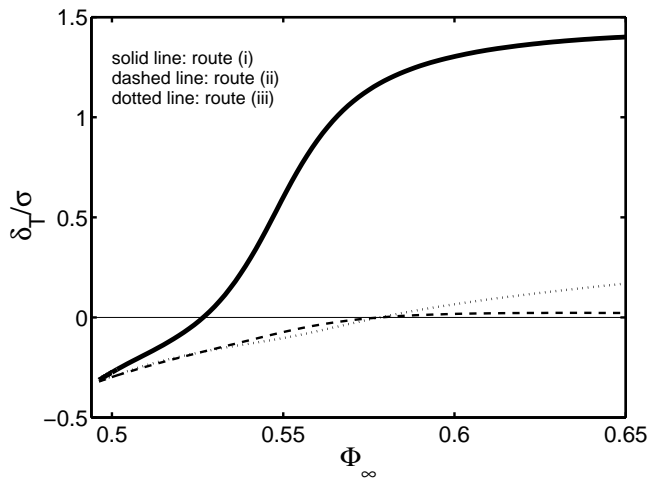}
\caption{Tolman length as a function of initial volume fraction
calculated along routes (i)--(iii).
}
\end{figure}

\section{Summary}

We presented a phase field theory of homogeneous crystal nucleation 
in hard sphere liquid, that describes local state of matter via a 
structural order parameter field and the density field. 

A rigorous test of the model is performed using the virtually exact
thermodynamic properties of the bulk phases. The two model parameters
are chosen so that the free energy and the $10\%-90\%$ thickness for
the cross-interfacial variation of the density peak height known
from molecular dynamics simulations are recovered. For the (111) and
(100) interfaces, and to a lesser extent for the (110) interface, the
resulting structural order parameter and density profiles are in a
significantly better agreement with the computer simulations,
than seen for previous continuum models. {\it Without adjustable
parameters}, the phase field theory predicts the height of the
nucleation barrier with a reasonable accuracy. This represents a
considerable improvement over the sharp interface droplet model of
the classical nucleation theory, which is known to overestimate the
nucleation rate by several orders of magnitude.

The cross-interfacial density profiles are used to calculate the Tolman
length for critical fluctuations, which is positive
for small sizes, and tends to a negative value in the large cluster limit.

\acknowledgments{
We thank R.~Davidchack and B.~B.~Laird for communicating us their
new high resolution density
data for the (111), (110), and (100) interfaces prior to publication
and for the illuminating
discussions. This work has been supported by the Norwegian Research Council
under project Nos.~153213/432 and~151400/210, by the ESA under Prodex
Contract No.~14613/00/NL/SFe, and by the Hungarian Academy of Sciences
under contract No.~OTKA-T-037323.
}


\begin{thebibliography}{99}
\bibitem{DEF}
  Tolman's original definition. A different definition is applied by 
  several authors: $\delta_T = lim_{R \rightarrow \infty} R_e - R_p$. 
  In our terminology, the latter corresponds to the equilibrium Tolman
  length, $\delta_{T,eq.}$.
\bibitem{DEF1}
  Defined by the position of a step-function that has the same spatial 
  integral and amplitude as the interfacial density profile.
\bibitem{DEF2}
  The surface at which the surface tension acts. It is the surface for 
  which the generalized Laplace equation, $\Delta p_c = 2\gamma/R +
  \partial \gamma/\partial R$ reduces to the original one; i.e., 
  $\partial \gamma/\partial R = 0$. Here $\Delta p_c$ is the capillary
  pressure.
\end{thebibliography}
\end{document}